\documentclass[final,3p,times]{elsarticle}
\usepackage{graphicx}  
\usepackage{bm}        
\usepackage{amsmath}
\usepackage{amssymb}
\usepackage[colorlinks,linkcolor=blue,citecolor=blue,urlcolor=blue]{hyperref}
\usepackage[all]{hypcap}
\usepackage{bbold}
\usepackage{subfigure}
\usepackage[title]{appendix}
\begin{document}
\begin{frontmatter}
\title{Spatial transient behavior in waveguides with lossy impedance boundary conditions}

\author[ad1,ad2]{Wei Guo}
\author[ad1,ad2]{Juan Liu}
\author[ad2]{Wenping Bi\corref{cor1}}
\ead{wenping.bi@univ-lemans.fr}
\author[ad1,ad3,ad4]{Desen Yang}
\author[ad2]{Yves Aur\'{e}gan}
\author[ad2]{Vincent Pagneux}
\address[ad1]{College of Underwater Acoustic Engineering, Harbin Engineering University, Harbin 150001, China}
\address[ad2]{Laboratoire d'Acoustique de l'Universit\'{e} du Mans, UMR-CNRS 6613, Le Mans, France}
\address[ad3] {Acoustic Science and Technology Laboratory, Harbin Engineering University, Harbin 150001, China}
\address[ad4]{Key Laboratory of Marine Information Acquisition and Security (Harbin Engineering University), Ministry of Industry and Information Technology, Harbin 150001, China}
\cortext[cor1]{Corresponding author. Laboratoire d'Acoustique de l'Universit\'{e} du Mans, UMR-CNRS 6613, Le Mans, France}

\begin{abstract}

Attenuation in acoustic waveguides with lossy impedance boundary conditions are associated with non-Hermitian and non-normal operators. This subject has been extensively studied in fundamental and engineering research, and it has been traditionally assumed that the attenuation behavior of total sound power can be totally captured by considering the decay of each transverse mode individually. One of the classical tools in this context  is the Cremer optimum concept that aims to maximize the attenuation of the least attenuated mode.
 However, a typical sound field may be a superposition of a large number of transverse modes which are nonorthogonal, and the individual mode attenuation may have little to do with the total sound power attenuation. By using singular value decomposition, we link the least attenuated total sound power to the maximum singular value of the non-normal propagator. 
 The behaviors of the least attenuated total sound power depend only on the lossy boundary conditions and frequency, but are independent of sources. 
 The sound may be almost non-decaying along the waveguide transition region for any lossy impedance boundary conditions although all modes attenuate exponentially.
This spatial transient appears particularly strongly if the impedance is close to an exceptional point of the propagator, at which a pair of adjacent modes achieve maximum attenuation predicted by Cremer optimum concept. These results are confirmed using non-modal numerical calculations and a two-by-two toy model.
\end{abstract}

\begin{keyword}
Spatial transient behavior, non-normality, lined duct, non-Hermitian, least attenuated sound power, Exceptional points, Cremer optimum impedance
\end{keyword}
\end{frontmatter}

\section{\label{sec1}Introduction}
Attenuation in acoustic waveguides with lossy impedance at the walls has been extensively investigated since the pioneering work of Morse~\cite{Morse1939} because it has important applications such as noise mitigation in aircraft engines or duct facilities~\cite{Nayfeh1975}. 
As in any waveguides, the sound pressure can be decomposed into a set of transverse modes, and, to maximize the global attenuation,
 the classical approach is to focus only on attenuation of each transverse mode individually. Following this line of thought, one of the fundamental results  is Cremer optimum concept~\cite{cremer} that aims to maximize the attenuation of the mode able to propagate on the longest distance (i.e. the least attenuated transverse mode).
According to this Cremer criterion, it appears that the maximum attenuation occurs when the wavenumbers of the two lowest transverse modes coalesce. Cremer optimum criterion has been one of the most important tool for duct liner design~\cite{Nayfeh1975, tester,  Mechel1991, Mechel1998, Zorumski1974, eversman1972, Eversman_book, Koch1977}, and it has been revisited recently in references~\cite{zhang2019}\cite{qiu2019}. 
It is true that if a source is dominated by a single transverse mode, considering the attenuation of each mode individually may describe the Total Sound Power (TSP) attenuation. However, typical sources are not so simple; usually, there is a mix of transverse modes with energy distributed among them in a manner that is generally unknown, and several studies  have resulted in the observation that Cremer criterion can be inefficient if an arbitrary source is considered~\cite{rice1968,Watson,Eversman_JSV_2007}.
The reason of this difficulty is that the transverse modes are non-orthogonal due to the lossy boundary condition at the walls, and a typical sound field is a superposition of a large number of transverse modes. Therefore, the sum of power attenuations of individual mode may have little to do with the TSP attenuation, meaning that the superposition of modes may dominate.

A matrix or operator whose eigenfunctions are non-orthogonal is said to be non-normal, and its eigenvalues may not totally capture its behavior. This can be illustrated by the failure of eigenvalue analysis in the hydrodynamic transition from a laminar to turbulent flow of Poiseuille and Couette flow~\cite{trefethen2005}. The critical Reynolds numbers (where the transition should occur) predicted by eigenvalue analysis are very different from those measured in the laboratories. Trefethen and  Embree~\cite{trefethen2005}, Trefethen~\textit{et al.}\cite{trefethen1993}, Reddy\textit{et al.}~\cite{reddy1993a}, and Butler and Farrel~\cite{butler1992} found that the transitions are induced by the time transient amplification of certain small perturbations although all modes decay monotonically along flow direction. 
This transient behavior appeared to be due to the non-normality of the operator governing the time evolution of the perturbations. Recently, similar transient behavior was found in a non-normal optical systems with an unbalanced distribution of loss and gain~\cite{makris2014}.  

In general, this non-normal transient behavior is closely connected to Exceptional Points (EPs) that have received much attention in recent years for non-Hermitian systems. 
The experimental demonstration of EPs has been  observed by Dembowski \textit{et al.}\cite{heiss2001} in a microwave cavity with dissipation. The important properties of EPs have been uncovered by Heiss\cite{heiss1990, heiss1991, heiss2004, heiss2012}, Rotter\cite{rotter2009}, and Berry\cite{berry2004} \textit{et al.}. EPs have been found in different systems, such as, laser-induced ionization states of atoms \cite{latinne1995}, electronic circuits \cite{Stehmann2004}, a chaotic optical microcavity\cite{lee2009}, PT-symmetric waveguides\cite{klaiman2009}, and waveguides with lossy impedance boundary conditions~\cite{newinsight}\cite{perrey2018}\cite{Midya2016} \textit{etc}. A range of extraordinary phenomena related with EPs have been illustrated, such as, loss-induced transparency\cite{guo2009}, enhancing the sensitivity in applications of microcavity sensors\cite{wiersig}, parametric instability\cite{zyablovsky}.

In this paper, we study the spatial transient behavior of the least attenuated TSP $G_{max}(x)$ as defined in Sec.~\ref{LTSP}, where $x$ is the axis coordinate, of a uniform waveguide governed by Helmholtz equation and lossy impedance boundary conditions which is described by the matrix operator $\mathsf{L}$ as defined in Eq. (\ref{eq_matrix_x}). We use singular value decomposition to link $G_{max}(x)$ to the maximum singular value of the propagator $\text{e}^{\text{i}x{\mathsf L}}$. We find that $G_{max}(x)$ shows almost no decay in transient region although all modes decay exponentially. By comparing the least attenuated TSP $G_{max}(x)$ with the attenuation of the least attenuated mode which dominates the modal attenuations, we find that only when $x\rightarrow\infty$, the least attenuated mode has the same decay rate with $G_{max}(x)$. In spatial transient region, they are different. This difference tend to be infinite when the admittance is close to the EPs at which a pair of modes achieve maximum attenuation predicted by Cremer optimum criterion. For each $x$, we find a least attenuated source that is generally close to a Gaussian form. When this source is imposed, the sound pressure field can take a pattern that avoids the lossy impedance wall and the TSP transports almost without decay in transient region, although all modes decay exponentially. This almost non-decayed TSP can be realized for any complex admittance, even in the vicinity of the EPs. A two-by-two toy model is also proposed to give in-depth understandings of these non-normality effects.
   
\section{Model}\label{model}

\begin{figure}[htbp]\label{fig2}
\centering\includegraphics[width=0.6\textwidth]{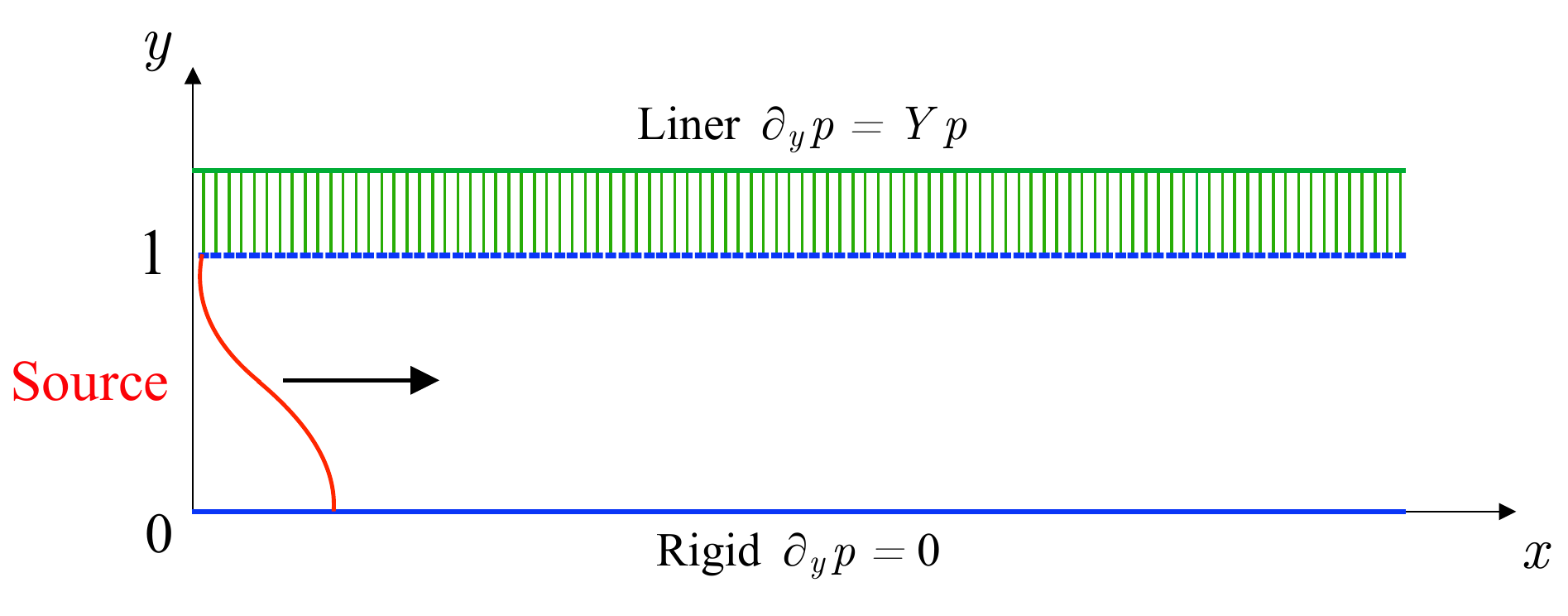}
\caption{(color online) Schematic depiction of a semi-infinite, two-dimensional waveguide with rigid boundary at $y=0$ and impedance boundary at $y=1$. A source is located at $x=0$.}
\end{figure}

We consider a semi-infinite, two-dimensional ($2$D) waveguide with one wall uniformly lined with locally reacting materials and the other wall acoustically rigid. Linear and lossless wave propagation is assumed. The configuration is depicted in Fig.~\ref{fig2}. The sound pressure satisfies the Helmholtz equation
\begin{equation}\label{eq1}
\frac{\partial^2 p}{\partial x^2}+\frac{\partial^2 p}{\partial y^2}+k^2p=0,
\end{equation}
with the boundary conditions
\begin{equation}\label{BC}
\frac{\partial p}{\partial y}\Bigm\vert_{y=1}=Y p,\; \;\;\;  \frac{\partial p}{\partial y}\Bigm\vert_{y=0}=0,
\end{equation}
where $Y=ik/Z$ and $Z$ refers to dimensionless impedance, it is a complex number. All lengths are divided by the waveguide width $h$. $k=\omega h/c$, where $\omega$ and $c$ refer to angular frequency and the sound velocity, respectively. Time dependence $\text{e}^{-\text{i}\omega t}$ is omitted in the followings. An acoustic source is located at $x=0$. The sound wave propagates along $x$ positive direction without reflection. 

We use a spectral collocation method based on Chebyshev polynomials to discretize sound pressure $p(x,y)$ along $y$, $\mathbf{p}^T=[p_1, p_2, \cdots, p_n, \cdots, p_N]^T=[p(y_1), p(y_2), \cdots, p(y_n), \cdots, p(y_N)]^T$, where $\mathbf{p}$ is a column vector, $``{}^T"$ refers to the transpose, $N$ refers to truncation number, the interpolation points $y_n=\cos[(n-1)\pi/(N-1)]$, $n=1,\;\cdots\; N$. Note that  $\mathbf{p}$ still depends on $x$ which has been omitted. The differential operator $\partial^2/\partial y^2$ is approximated by $\mathsf{D_2}$ calculated by the MATLAB program \textit{chebdif}. By taking only the right-going wave, Eq. (\ref{eq1}) is written as 
\begin{equation}\label{eq_matrix_x}
\frac{d\mathbf{p}}{dx}=\text{i}\mathsf{L}\mathbf{p},
\end{equation}
where $\mathsf{L}=\sqrt{\mathsf{D_2}+k^2\mathsf{I}}$ is a matrix operator, $\mathsf{I}$ refers to identity matrix. A detailed derivation of Eq. (\ref{eq_matrix_x}) is shown in Appendix~\ref{AppendixI}. It needs to stress that the boundary conditions (\ref{BC}) have been included in the matrix operator $\mathsf{L}$. $\mathsf{L}$ is non-normal, i.e., $\mathsf{L}\mathsf{L}^\dagger\ne\mathsf{L}^\dagger\mathsf{L}$ when $Y$ is complex, where $``{}^\dagger"$ refers to the complex conjugate transpose. The modes of this operator are not orthogonal. $\mathsf{L}$ is also non-Hermitian, i.e., $\mathsf{L}\ne\mathsf{L}^\dagger$. For a non-Hermitian operator, the eigenvalues and the corresponding eigenfunctions of a pair of adjacent modes may coalesce at some special points in the complex parameter plane, e.g., admittance plane in our case. These points are called Exceptional points (EPs) as we have pointed out in the Introduction. In our case, there are infinite number of EPs~\cite{newinsight} in the complex value admittance plane. In this work, we consider only the first one. 

The solution to Eq. (\ref{eq_matrix_x}) is 
\begin{equation}\label{solution_vec}
	 \mathbf{p}(x)=\text{e}^{\text{i}x\mathsf{L}}\mathbf{p}(0).
\end{equation}
The TSP can then be expressed in terms of sound pressure distribution vector $\mathbf{p}$ 
\begin{equation}\label{eq_p2_vec}
\int_0^1|p(x,y)|^2\, \mathrm{d}y=\mathbf{p}^\dagger \mathsf{W}\mathbf{p}=(\mathsf{F}\mathbf{p} )^\dagger\mathsf{F}\mathbf{p}=\|\mathsf{F}\mathbf{p}\|^2=\|\mathbf{q}(x)\|^2,
\end{equation}
where we have used the Clenshaw-Curtis quadrature to numerically calculate the integral $\int_0^1|p(x,y)|^2\, \mathrm{d}y$, and for simplicity, we have omitted the independent variable $x$ for vector function $\mathbf{p}(x)$. $\mathsf{W}$ is the weight matrix, it is diagonal and positive definite. It can be decomposed into $\mathsf{W}=\mathsf{F}^\dagger\mathsf{F}$ using a Cholesky decomposition, where $\mathsf{F}$ is still a diagonal matrix. In the last equation, we have assumed $\mathbf{q}=\mathsf{F}\mathbf{p}$ and $\|\mathbf{q}\|$ refers to $L_2$-norm of the complex vector $\mathbf{q}$.

\section{\label{LTSP}The least attenuated TSP}

We define the attenuation of TSP $G(x)$ as
\begin{equation}\label{eq_def_power}
 G(x)=\frac{\int_0^1{|p(x,y)|^2}\, \mathrm{d}y}{\int_0^1{|p(0,y)|^2}\, \mathrm{d}y},
\end{equation}
where the TSP at $x$ is defined as $\int_0^1|p(x,y)|^2\, \mathrm{d}y=\int_0^1p^*(x,y)p(x,y)\, \mathrm{d}y$. $G(x)$ is different from the usual definition of energy flux attenuation in waveguides. We will show in Sec.~\ref{sec_flux} that this difference does not change our conclusions. Substituting Eq. (\ref{eq_p2_vec}) into the definition (\ref{eq_def_power}), the attenuation of TSP $G(x)$ can be calculated as
\begin{equation}\label{G_q}
G(x)=\frac{\|\mathbf{q}(x)\|^2}{\|\mathbf{q}_0\|^2},
\end{equation}
where $\mathbf{q}_0=\mathsf{F}\mathbf{p}_0$ represents sources. In the whole paper, $\mathbf{q}_0$ is normalized so that $\|\mathbf{q}_0\|^2=\mathbf{q}^{\dagger}_0\mathbf{q}_0=\int_0^1|p(0,y)|^2\mathrm{d}y=1$, i.e., we assume the sound power of the source is normalised to $1$. The variations of $G(x)$ along $x$ depend strongly on the sources~\cite{Eversman_JSV_2007}. 

The least attenuated TSP, at each $x$, can be achieved by ``optimizing" $G(x)$ over all permissible sources~\cite{schmid2007},
\begin{eqnarray}\label{least_TSP}
G_{max}(x) &=& \max\limits_{\mathbf{q}_0}G(x)
=\max\limits_{\mathbf{q}_0}\frac{\|\mathbf{q}(x)\|^2}{\|\mathbf{q}_0\|^2}
=\max\limits_{\mathbf{q}(0)}\frac{\|\mathsf{F}\text{e}^{\text{i}x\mathsf{L}}\mathsf{F}^{-1}\mathbf{q}_0\|^2}{\|\mathbf{q}_0\|^2}\nonumber\\
&=&\|\mathsf{F}\text{e}^{\text{i}x\mathsf{L}}\mathsf{F}^{-1}\|^2
=\max(\sigma_n)^2,
\end{eqnarray}
where we have used $\mathbf{q}(x)=\mathsf{F}\text{e}^{\text{i}x\mathsf{L}}\mathsf{F}^{-1}\mathbf{q}_0$ and invoked the definition of a vector-induced matrix norm in above derivation. $\sigma_n$ are the singular values of matrix $\mathsf{F}\text{e}^{\text{i}x\mathsf{L}}\mathsf{F}^{-1}$. 

\begin{figure}\label{fig_envelop}
\centering\includegraphics[width=0.5\textwidth]{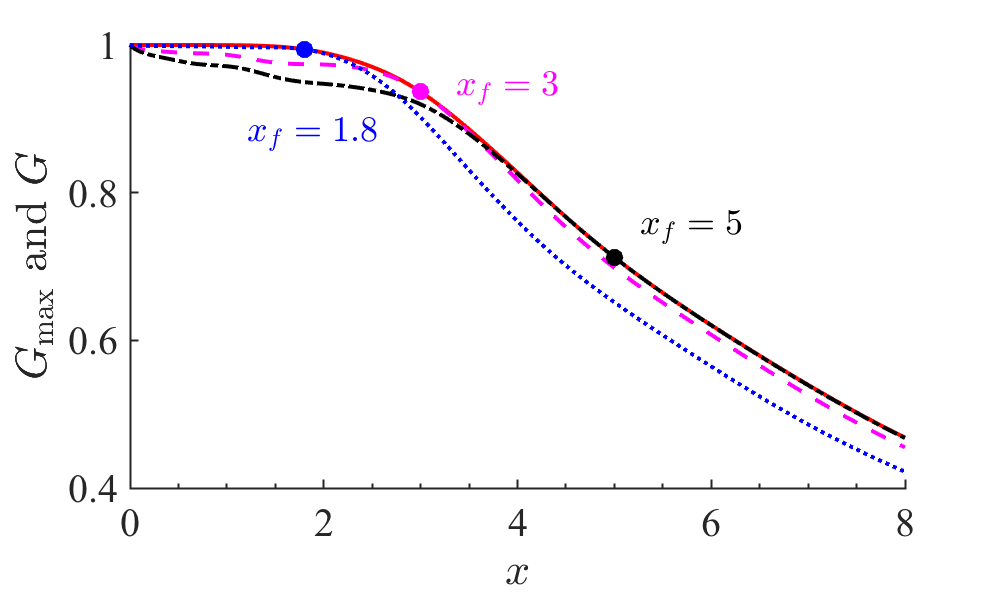}
\caption{(color online) Curve $G_{max}(x)$ is an envelop of $G(x)$. The TSP attenuation $G(x)$ are calculated by three different least attenuated sources $\mathbf{q}_{least}(x_f)$ in order to obtain maximum TSP at $x_f=1.8$ (dot line), $x_f=3$ (dashed line), and $x_f=5$ (dot-dashed line), respectively. $k=10$ and $Y=1+4\text{i}$.}
\end{figure}

\textit{For each given $x$, $G_{max}(x)$ in Eq. (\ref{least_TSP}) gives the least attenuated TSP by taking care of the optimization over all permissible sources for a given waveguide with lossy impedance boundary conditions and frequency which is described by matrix operator $\mathsf{L}$.  $G_{max}(x)$ is totally decided by $exp(\text{i}x\mathsf{L})$ (Note that $\mathsf{F}$ is an integration weight matrix.), therefore it is independent of sources.} However, it is important to realize that, for each given $x=x_f$, the least attenuated TSP $G_{max}(x_f)$ is achieved by using a different source which is given by the principal right singular vector of matrix operator $\mathsf{F}\text{e}^{\text{i}x\mathsf{L}}\mathsf{F}^{-1}$~\cite{trefethen2005}~\cite{schmid2007}. We call this source the least attenuated source, expressed as $\mathbf{q}_{least}(x_f)$. When this source inputs, the usually calculated TSP attenuation $G(x)$ as defined in Eq. (\ref{G_q}) will be maximum at $x=x_f$ by taking care of the optimization over all permissible sources, and $G_{max}=G$ at $x=x_f$ as shown in Fig.~\ref{fig_envelop} when $x_f=1.8$, $x_f=3$, and $x_f=5$. Therefore, curve $G_{max}(x)$ is not a TSP attenuation curve $G(x)$ as usually illustrated in the literature of lined ducts, but an envelop of $G(x)$. From the point of view of noise reduction in lined ducts, $G_{max}(x)$ gives the worst case at each $x$.

$G_{max}(x)$ can also be interpreted in term of modes. In Eq. (\ref{least_TSP}), using eigenvalue decomposition of matrix, $\mathsf{L}=\mathsf{V}\mathsf{\Lambda}\mathsf{V}^{-1}$, where $\mathsf{V}$ is a matrix with eigenvectors as columns, $\Lambda$ is a diagonal matrix with eigenvalues on the main diagonal. Eq. (\ref{least_TSP}) is then rewritten as
\begin{equation}\label{G_mode}
G_{max}(x)=\|\mathsf{F}\text{e}^{\text{i}x\mathsf{L}}\mathsf{F}^{-1}\|^2=\|\mathsf{F}\mathsf{V}\text{e}^{\text{i}x\mathsf{\Lambda}}\mathsf{V}^{-1}\mathsf{F}^{-1}\|^2.
\end{equation}
$\|\text{exp}(\text{i}x\mathsf{\Lambda})\|^2$ represents the contributions of eigenvalues. If the boundary conditions are lossy, it represents mode attenuations and is dominated by the attenuation of the least attenuated mode,
\begin{equation}
M_{max}(x)=e^{-2\text{Im}(k_1)x},
\end{equation}
where Im$(k_1)$ denotes the imaginary part of the eigenvalue of the least attenuated mode. Only when the boundary conditions are non-lossy ($Y$ is a pure real number), is $\mathsf{L}$ normal, $\mathsf{V}$ a unitary matrix,  $\|\mathsf{V}\|^2\|\mathsf{V}^{-1}\|^2=1$, and $G_{max}(x)$ totally decided by $\|\text{exp}(\text{i}x\mathsf{\Lambda})\|^2$. Otherwise $\mathsf{L}$ is non-normal, $\mathsf{V}$ is not a unitary, $\|\mathsf{V}\|^2\|\mathsf{V}^{-1}\|^2$ may be very large when $\mathsf{L}$ is far from normality~\cite{schmid2007}. $G_{max}(x)$ may be highly different from $M_{max}(x)$ when $\mathsf{L}$ is far from normality~\cite{trefethen2005}. $M_{max}(x)$ cannot capture the behavior of $G_{max}(x)$. Therefore, we define 
\begin{equation}\label{R_eq}
R(x)=\frac{G_{max}(x)}{M_{max}(x)},
\end{equation}  
to describe the departure of the least attenuated TSP $G_{max}(x)$ from the attenuation of the least attenuated mode $M_{max}(x)$. The farther away from normality of the matrix operator $\mathsf{L}$ is, the larger is the departure of $G_{max}(x)$ from $M_{max}(x)$.
\begin{figure}\label{fig_Gmax}
\centering\includegraphics[width=0.8\textwidth]{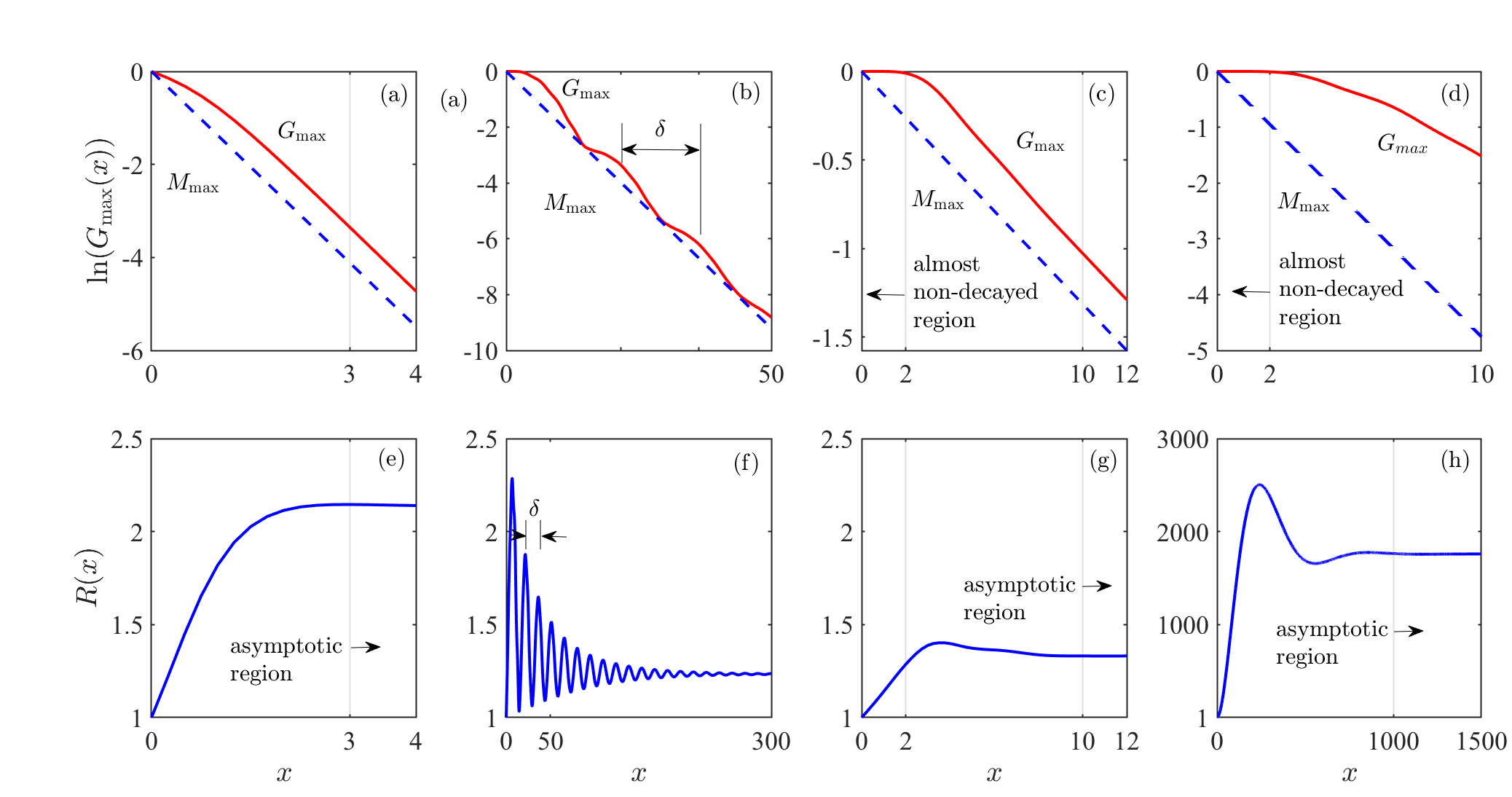}
\caption{(color online) The least attenuated TSP $G_{max}(x)$ and $R(x)$ the departure of $G_{max}(x)$ from $M_{max}(x)$. The attenuations of the least attenuated modes $M_{max}(x)$ are also plotted for comparison.  (a) and (e) $k=2$, $Y=1+2\text{i}$. (b) and (f) $k=10$, $Y=1+\text{i}$. (c) and (g) $k=10$, $Y=1+4\text{i}$. (d) and (h) $k=10$, $Y=1.65+2.06\text{i}$ is close to EP.}
\end{figure}

Typical $G_{max}(x)$ and $R(x)$ as functions of $x$ are plotted in Fig.~\ref{fig_Gmax} when $k=2$ for $Y=1+2\text{i}$ ((a) and (e)) and $k=10$ for $Y=1+\text{i}$ ((b) and (f)), $Y=1+4\text{i}$ ((c) and (g)), and $Y=1.65+2.06\text{i}$ which is close to EP ((d) and (h)), respectively. For comparison, we also plot the attenuations of the least attenuated modes $M_{max}(x)$ in Fig.~\ref{fig_Gmax} (a)-(d). 

$R(x)$ tend to be constant when $x$ goes beyond certain values $x_{\text{tran}}$, e.g., $x_{\text{tran}}\approx 3$ in (e), $x_{\text{tran}}\approx 200$ in (f), $x_{\text{tran}}\approx 10$ in (g), and $x_{\text{tran}}\approx 1000$ in (h). This means that $G_{max}(x)$ decay exponentially at the same rate as $M_{max}(x)$ as shown, e.g., in (a) and (c). This is the asymptotic region. Only in this region, $M_{max}(x)$ dominates the TSP and capture the decayed rates of $G_{max}(x)$. However, $G_{max}(x)$ and $M_{max}(x)$ still differ by a constant. This constant is a function of $Y$,  and when $Y$ is close to the EP as shown in Fig.~\ref{fig_Gmax} (h), this constant becomes large. 

When $x<x_{\text{tran}}$, $G_{max}(x)$ and $R(x)$ are in spatial transient region. In Fig.~\ref{fig_Gmax} (a), $k=2$ and $Y=1+2$i, for this very low frequency, it is often assumed that higher modes attenuate rapidly, the least attenuated mode dominate the TSP. However, it is easy to see that the difference between $G_{max}(x)$ and $M_{max}(x)$ is not negligible. When $Y$ is close to the EP (not shown in this figure), the difference becomes large.

When $k$ increases, transient region can be roughly divided into two sub-regions: (1) almost non-decayed transient regions in which $G_{max}(x)$ is almost non-decayed although all modes attenuate exponentially. This region is approximately in $0<x<2$ for $k=10$ as shown in Fig.~\ref{fig_Gmax} (c) and (d). This almost non-decayed transient region can be realized for any $Y$, even when $Y$ is close to the EPs, as shown in Fig.~\ref{fig_Gmax} (d) for $Y=1.65+2.06$i, at which Cremer optimum criterion predicts that a pair of modes achieve maximum attenuation. (2) damped-oscillation transient region in which the behavior of $G_{max}(x)$ may be analogous to that of the displacement of a damped harmonic oscillator. In this region, $G_{max}(x)$ may be decayed oscillation as shown in Fig.~\ref{fig_Gmax} (b) when it is ``underdamping" or decay non-exponentially as shown in Fig.~\ref{fig_Gmax} (c) and (d) when it is ``overdamping". The analogy of ``underdamping" and ``overdamping"  can be seen more clearly in Fig.~\ref{fig_Gmax} (f)-(h) for $R(x)$. As will be shown by a two-by-two toy model in Sec.~\ref{sec:toy}, the periods of the decayed oscillations $\delta$ are decided by the difference of real parts of the axial wavenumbers. When $k=10$, the range of damped-oscillation transient regions start roughly at $x=2$ as shown in Fig.~\ref{fig_Gmax} (b)-(d) and stop roughly at $x=x_{\text{tran}}$ as shown in Fig.~\ref{fig_Gmax} (f)-(h).

\section{\label{indicator}Non-normality indicator}

In Fig.~\hyperref[fig_Gmax]{\ref{fig_Gmax}(e), (f), (g), and (h)} we observe that $R(x)$, which describes the departure of $G_{max}$ from $M_{max}$, always start at $1$ and then grows up to its maximum. We use this maximum departure to define an indicator of non-normality for the matrix operator $\mathsf{L}$
\begin{figure}\label{fig_H_beta}
\centering
\includegraphics[width=0.8\textwidth]{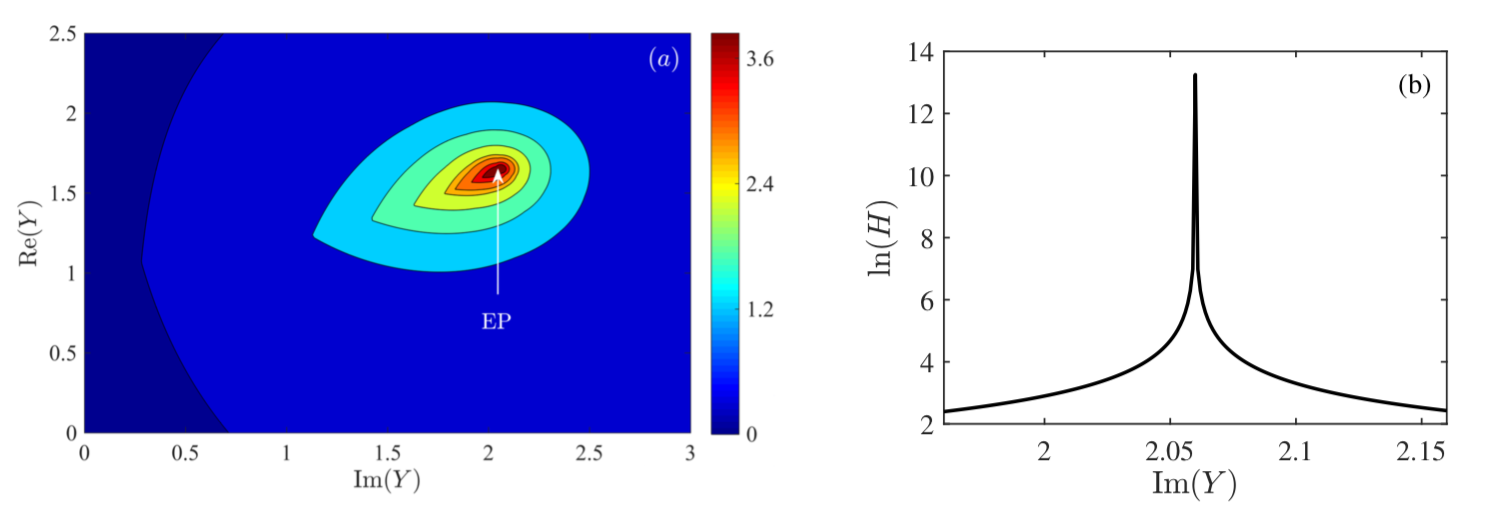}
\caption{Contour plot of non-normality indicator ln($H$) in the complex $Y$ plane (left panel) and ln($H$) as a function of $\text{Im}(Y)$ when $\text{Re}(Y)=\text{Re}(Y_{\text{EP}})$ in the vicinity of the first EP (right panel). $Y_{\text{EP}}=1.650611293539765+2.059981457179885\text{i}$, $k=10$.}
\end{figure}
\begin{equation}\label{eq12}
H=\max\limits_x R(x)=\max\limits_x\frac{G_{\max}(x)}{M_{\max}(x)}.
\end{equation}
$H$ is a function of $Y$.

We plot the contours of non-normality indicator ln($H$) in the complex $Y$ plane as shown in the left panel of Fig.~\hyperref[fig_H_beta]{\ref{fig_H_beta}} for $k=10$. It is shown that $H=1$ is minimum, it is located on axis of Im$(Y)=0$, corresponding to $\mathsf{L}$ is normal. $1\le H<3$ for most $Y$. $H$ increases when $Y\rightarrow Y_{\text{EP}}=1.650611293539765+2.059981457179885\text{i}$. The increasing behavior of $H$ as a function of Im($Y$) when $\text{Re}(Y)=\text{Re}(Y_{\text{EP}_1})$ is shown in Fig.~\ref{fig_H_beta}. We find numerically that $H\propto 1/|Y-Y_{\text{EP}}|$. We will return to this point in Sec.~\ref{sec:toy}.

\section{\label{sec_source}The least attenuated sources and the corresponding sound fields}

As we have indicated in Sec.~\ref{LTSP}, for each given $x=x_f$, the least attenuated TSP $G_{max}(x_f)$ is achieved by using a least attenuated source $\mathbf{q}_{least}(x_f)$. When this source inputs, the usually calculated TSP attenuation $G(x)$ as defined in Eq. (\ref{G_q}) will be maximum at $x=x_f$ by taking care of the optimization over all permissible sources. $\mathbf{q}_{least}(x_f)$ is the principal right singular vector of matrix operator $\mathsf{F}\text{e}^{\text{i}x\mathsf{L}}\mathsf{F}^{-1}$~\cite{schmid2007}. We find that when $x$ in the almost non-decayed transient region, the amplitude $|\mathbf{q}_{{least}}|$ is approximately a half of Gaussian function along $y$ (If the boundary condition at $y=0$ is same as that of $y=1$, $|\mathbf{q}_{{least}}|$ will be approximately a  Gaussian function.). 
\begin{figure}[ht]\label{fig_source}
\centering
\includegraphics[width=0.6\textwidth]{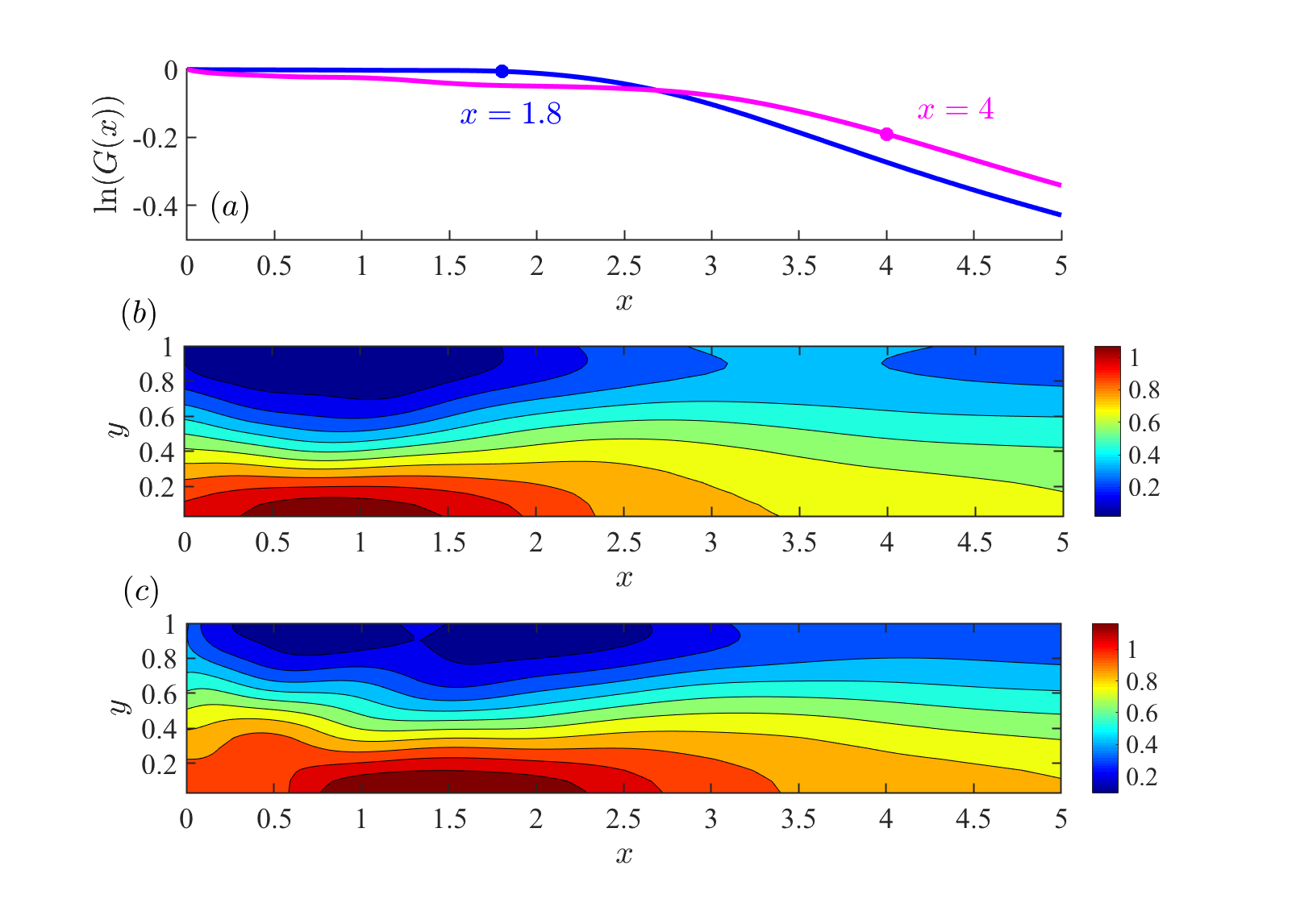}
\caption{(color online) (a) The attenuations of TSP ln($G$) as a function of $x$ when the least attenuated sources $\mathbf{q}_{{least}}(x_f=1.8)$ and $\mathbf{q}_{{least}}(x_f=4)$ are incident, respectively. (b) The corresponding sound pressure field $|p(x,y)|$. $Y=1+4\text{i}$ and $k=10$.}
\end{figure}

In Fig.~\ref{fig_source} (a), we plot the TSP attenuation ln($G(x)$) when the least attenuated sources $\mathbf{q}_{least}(x_f=1.8)$ and $\mathbf{q}_{least}(x_f=4)$ are incident, respectively, for $Y=1+4\text{i}$ and $k=10$. $x=1.8$ is still in the almost non-decayed transient region ($x<2$) as shown in Fig. ~\ref{fig_Gmax} (c) and (g). The TSP transports almost non-decayed in the range of $x<1.8$. The corresponding sound pressure $|p(x,y)|$ is shown in Fig. ~\ref{fig_source} (b). It is highly interesting that the sound field can avoid the lossy impedance boundary, like a half of a Gaussian beam, in the almost non-decayed transient region. However, $x=4$ is out of almost non-decayed transient region as shown in Fig.~\ref{fig_Gmax} (c) and (g). Although $G(x)=G_{max}(x)$ at $x=4$ is still maximum by considering all possible sources, $\mathbf{q}_{{least}}(x_f=4)$ deviates from Gaussian function; $G$ decays in the transient region; $|p(x,y)|$ gradually increases in the vicinity of the lossy impedance boundary as shown in Fig.~\ref{fig_source} (c). 

We stress that the almost non-attenuated TSP in transient region can be realized for any $Y$, even in the vicinity of the EPs at which the the rates of exponential attenuations of a pair of modes which dominate the TSP are maximum. 

\section{\label{sec:toy}Toy model}

In this section, we construct a two-by-two toy model and analytically analyze the behavior of $G_{\max}$, $R(x)$, and $H$. To solve Eq. (\ref{eq1}) with the boundary conditions (\ref{BC}), we expand the sound pressure as proposed in Ref.~\cite{modelling06}
\begin{equation}\label{eq_expand}
p(x,y)=\sum_{n=0}^Np_n(x)\psi_n(y),
\end{equation}
where $\psi_n(y)$ are the normalized eigenfunctions of rigid modes which satisfy the Helmholtz equation (\ref{eq1}) and rigid boundary conditions $\partial p/\partial y=0$ at $y=0$ and $y=1$ respectively, $N$ refers to the truncation number. Using Eq. (\ref{eq_expand}), we project Eqs. (\ref{eq1}) and (\ref{BC}) onto the corresponding rigid modes $\psi_n(y)$, we obtain 
\begin{equation}\label{toy1}
\frac{\mathrm{d}\mathbf{p}}{\mathrm{d}x}=\text{i}\mathsf{L}\mathbf{p},
\end{equation}
where
\begin{eqnarray}\label{toy_matrix}
\mathsf{L} &=& \sqrt{\mathsf{L}_1},\label{toy_L}\\
\mathsf{L}_1 & =& \begin{bmatrix}k^2-\alpha_0^2 & 0\\ 0 & k^2-\alpha_1^2\end{bmatrix}+Y\begin{bmatrix}1 & -\sqrt{2}\\ -\sqrt{2} & 2\end{bmatrix}=\begin{bmatrix}a_1+Y & -\sqrt{2}Y\\ -\sqrt{2}Y & a_2+2Y\end{bmatrix}.\label{toy_matrix}
\end{eqnarray}
In the derivations of above equations, we have chosen the truncation number $N=1$. $\alpha_0=0$ and $\alpha_1=\pi$ refer to transverse eigenvalues of the first two rigid modes. $\mathbf{p}=[p_0(x), p_1(x)]^T$, where $``{}^T"$ refers to the transpose. $a_1=k^2$ and $a_2=k^2-\alpha_1^2=k^2-\pi^2$ are real, therefore $a_1-a_2=\pi^2$. The detailed derivations of Eqs. (\ref{toy1}-\ref{toy_matrix}) are shown in Appendix~\ref{AppendixI}. Equation (\ref{toy1}) is similar to Eq. (\ref{eq_matrix_x}) except the dimensions of  $\mathbf{p}$ and $\mathsf{L}$.

The two eigenvalues $k_{n}=\sqrt{\gamma_{n}}$ of matrix $\mathsf{L}$ coalesce, where $\gamma_n$ refer to the eigenvalues of matrix $\mathsf{L}_1$, $n=1,\; 2$, when the parameter $Y=Y_{\text{EP}}$ (or $Y_{\text{EP}}^*$), where 
\begin{equation}
Y_{\text{EP}}=\frac{\pi^2}{9}\left(1+2\sqrt{2}i\right).
\end{equation}
From Eq. (\ref{eigv_x}), the corresponding eigenvectors also coalesce. Therefore $Y_{\text{EP}}$ and $Y_{\text{EP}}^*$ are two EPs. By keeping in mind that Im($Y$) must be non-negative for a passive system, we take only $Y_{\text{EP}}$ for our lossy waveguide problem. Note that to assure the truncation of the expansion (\ref{eq_expand}) to $N=1$ represents a true passive sound pressure field in a waveguide with lossy impedance boundary conditions, we have chosen the imaginary parts of eigenvalues $k_{1,2}$ to be non-negative (convention $\exp(-\text{i}\omega t)$), and proven $dW(x)/dx<0$ as shown in Appendix~\ref{AppendixII}. 

To analytically calculate the least attenuated TSP $G_{max}$, we use $\mathbf{p}=\mathsf{T}\mathbf{q}$ to transform matrix $\mathsf{L}$ to a triangle matrix $\mathsf{L}'$, where $\mathsf{T}$ is a unitary matrix by which the transformation conserves the TSP. Eq. (\ref{toy1}) can be written as
\begin{equation}\label{toy11}
\frac{\mathrm{d}{\mathbf{q}}}{\mathrm{d}x}=\text{i}{\mathsf{L'}}\mathbf{q},\;\; \text{where},\;\; \mathsf{L}'=\mathsf{T}^{-1}\mathsf{L}\mathsf{T}=\begin{bmatrix}k_1 & C_1 \\ 0 & k_2 \end{bmatrix}.
\end{equation}
The detailed derivations of Eq. (\ref{toy11}), the elements of matrix $\mathsf{T}$, and $C_1$ are shown in Appendix~\ref{AppendixIII}. 

The solution of Eq. (\ref{toy11}) is 
\begin{equation}
\mathbf{q}(x)=\text{e}^{\text{i}x\mathsf{L}'}\mathbf{q}_0,
\end{equation}
where $\mathbf{q}_0$ refers to a source at $x=0$, the exponential of matrix $\mathsf{L}'$ is
\begin{equation}
e^{\text{i}x\mathsf{L}'}=\begin{bmatrix}e^{\text{i}k_1x} & \frac{e^{\text{i}k_1x}-e^{\text{i}k_2x}}{k_1-k_2}C_1\\ 0 & e^{\text{i}k_2x}\end{bmatrix}.  
\end{equation} 
Similar to Eq. (\ref{least_TSP}), the least attenuated TSP is
\begin{eqnarray}\label{tbt_Gmax}
G_{max} &=& \max\limits_{\mathbf{q}_0}\frac{\|\mathbf{q}(x)\|^2}{\|\mathbf{q}_0\|^2}=\max\limits_{\mathbf{q}_0}\frac{\|\text{e}^{\text{i}x\mathsf{L}'}\mathbf{q}_0\|^2}{\|\mathbf{q}_0\|^2}\nonumber\\
&=&\|\text{e}^{\text{i}x\mathsf{L}'}\|^2= \max(\sigma_n)^2\nonumber\\
&=& M_{max}(x)R(x),
\end{eqnarray}
where
\begin{equation}\label{tbt_MR}
M_{max}(x)=e^{-2\text{Im}(k_1)x},\;\; R(x)=\frac{1}{2}\left(\text{tr}+\sqrt{\text{tr}^2-4e^{-2k_{\mathbb{I}}x}}\right),
\end{equation}
$\sigma_n$ refer to the singular values of exponential matrix $e^{\text{i}x\mathsf{L}'}$, 
\begin{eqnarray}\label{G_analy1}\nonumber
\text{tr} &=& \left(1+\frac{\vert C_1\vert^2}{\vert k_1-k_2\vert^2}\right)(1+e^{-2k_{\mathbb{I}}x})-\frac{2\vert C_1\vert^2\cos(k_{\mathbb{R}} x)}{\vert k_1-k_2\vert^2}e^{-k_{\mathbb{I}}x}\\
            & =& 1+e^{-2k_{\mathbb{I}}x}+C_3\text{Im}(Y)^2\left[e^{-2k_{\mathbb{I}}x}-2\cos(k_\mathbb{R}x)e^{-k_{\mathbb{I}}x}+1\right]P_t,\\\nonumber
	k_{\mathbb{R}} &=& \text{Re}(k_2-k_1),\\\nonumber 
	k_{\mathbb{I}} &=&\text{Im}(k_2-k_1),\\\nonumber
	P_t &=& \frac{1}{\vert\bm{X}_1{^T}\bm{X}_1\vert^2}=\frac{4C_2}{\vert \gamma_1-\gamma_2\vert^2}=\frac{4C_2}{\vert Y-Y_{\text{EP}}\vert\vert Y-Y_{\text{EP}}^*\vert},
\end{eqnarray}
where $C_2 =\Lambda_1^4/\vert \pi^2-Y-(\gamma_1-\gamma_2)\vert^2$ and $C_3=\pi^4/2C_2^2$. In the expression of $P_t$, we have used
\begin{equation}\label{gamma_YEP}
\gamma_1-\gamma_2=\sqrt{(Y-Y_{\text{EP}})(Y-Y_{\text{EP}}^*)}.
\end{equation}
$P_t$ refers to Petermann factor of mode $1$. When $Y\rightarrow Y_{\text{EP}}$, $P_t\rightarrow\infty$. Petermann factor, which describes the self-overlap of a mode~\cite{newinsight}, plays important roles in excess-noise for unstable lasers~\cite{Petermann1979}. Recently, its important roles have also been found in open quantum systems~\cite{Lee2009PRA} and waveguides with lossy boundary conditions~\cite{newinsight}. Equations (\ref{tbt_Gmax}-\ref{G_analy1}) show clearly that $G_{max}$ is not dependent on the sources whose effects are described by $p_n$ in Eq. (\ref{eq_expand}). 

Note that, without loss of the generality, we have assumed mode $1$ to be the least attenuated mode in Eqs. (\ref{tbt_Gmax}) and (\ref{tbt_MR}), i.e., $\text{Im} (k_1)<\text{Im} (k_2)$. If, on the other hand,  $\text{Im} (k_2)<\text{Im} (k_1)$, $M_{max}(x)=e^{-2\text{Im}(k_2)x}$, the least attenuated mode is mode $2$. Note that in this section, we will analytically analyze the behavior of $R$ and $G_{max}$ as functions of $x$ and complex $Y$. We will use alternatively $x$ and $Y$ as independent variable.

We can find from Eqs. (\ref{tbt_MR}) and (\ref{G_analy1}) that three factors contribute to the departure of $G_{max}$ from $M_{max}$: $P_t\propto1/\vert Y-Y_{\text{EP}}\vert$ where $\vert Y-Y_{\text{EP}}\vert$ describes the departure from the EP in the complex $Y$ plane; $\cos(k_{\mathbb{R}}x)$ represents the interferences between the modes; and $\exp(-k_{\mathbb{I}}x)$ represents the differences of attenuation rates between the modes. $\cos(k_{\mathbb{R}}x)\exp(-k_{\mathbb{I}}x)$ in Eq. (\ref{G_analy1}) dominates the variation of $R(x)$ in most cases. It may be analogous to the displacement of a damped harmonic oscillator. The $Y$ plane could be roughly divided into two parts: in the region Im$(Y)<\text{Im}(Y_{\text{EP}})$, $k_{\mathbb{R}}=O(k_{\mathbb{I}})$ or $\vert k_{\mathbb{R}}\vert>\vert k_{\mathbb{I}}\vert$, the transient behavior of $R(x)$ is in damped oscillation as a underdamped oscillator as shown in Fig. ~\ref{fig_Gmax} (f); in the region Im$(Y)>\text{Im}(Y_{\text{EP}})$, $k_\mathbb{I}$ becomes larger as Im$(Y)$ increases. $R(x)$ decays as a overdamped oscillator as shown in Fig.~\ref{fig_Gmax} (g) and (h).

When $x$ is small such that $k_{\mathbb{R}}x\ll 1$ and $k_{\mathbb{I}}x\ll 1$
\begin{equation}\label{Rx_approx}
R(x)\approx 1+\left(\sqrt{k_\mathbb{I}^2+C_4\frac{k_\mathbb{R}^2+k_\mathbb{I}^2}{\vert\gamma_1-\gamma_2\vert^2}}-k_\mathbb{I}\right)x+\frac{1}{2}\left(2k_\mathbb{I}^2+C_4\frac{k_\mathbb{R}^2+k_\mathbb{I}^2}{\vert\gamma_1-\gamma_2\vert^2}-\frac{k_\mathbb{I}\left(2k_\mathbb{I}^2+C_4\frac{k_\mathbb{R}^2+k_\mathbb{I}^2}{\vert\gamma_1-\gamma_2\vert^2}\right)}{\sqrt{k_\mathbb{I}^2+C_4\frac{k_\mathbb{R}^2+k_\mathbb{I}^2}{\vert\gamma_1-\gamma_2\vert^2}}}\right)x^2,
\end{equation}
as derived in Appendix~\ref{AppendixIV}, where $C_4=4C_2C_3\text{Im}(Y)^2$ is a positive real number. The coefficients of $x$ and $x^2$ are positive as shown in Appendix~\ref{AppendixIV}. Therefore, $R(x)$ always increases when $x$ is small. 

For any given $Y$ and $Y\ne Y_{\text{EP}}$, when $x\rightarrow\infty$, $e^{-k_{\mathbb{I}}x}\rightarrow 0$, $R(x)$ turns to be a constant,
\begin{equation}\label{RxinftyYfix}
R(x)\rightarrow 1+C_3\text{Im}(Y)^2P_t,
\end{equation}
where $C_3$ is a positive real number. As $\text{Im}(Y)\rightarrow 0$, this constant is close to $1$. This constant will be very large when $Y$ is close to $Y_{\text{EP}}$ because of $P_t\propto1/\vert Y-Y_{\text{EP}}\vert$. $G_{max}(x)$ then tends to $[1+C_3\text{Im}(Y)^2P_t]M_{max}(x)$, decreases exponentially with the same rate as the least attenuated mode $M_{max}$.

For any given $x$, when $Y\rightarrow Y_{\text{EP}}$, we have $k_\mathbb{R}\rightarrow 0$, $k_\mathbb{I}\rightarrow 0$, $k_\mathbb{R}x\ll 1$, $k_\mathbb{I}x\ll 1$, and $P_t\rightarrow\infty$, the approximation of $R(Y)$ can be obtained as same as that given in Appendix~\ref{AppendixIV} for $x$ small, 
\begin{eqnarray}\label{R_const}
R(Y) &\approx& 1+\sqrt{C_4\frac{k_\mathbb{R}^2+k_\mathbb{I}^2}{\vert\gamma_1-\gamma_2\vert^2}}x+\frac{1}{2}\left(C_4\frac{k_\mathbb{R}^2+k_\mathbb{I}^2}{\vert\gamma_1-\gamma_2\vert^2}\right)x^2\nonumber\\
&\approx& 1+\sqrt{\frac{C_4}{4\vert a_1+a_2+3Y\vert}}x+\frac{C_4}{8\vert a_1+a_2+3Y\vert}x^2,
\end{eqnarray}
where the approximation of $(k_\mathbb{R}^2+k_\mathbb{I}^2)/(\vert\gamma_1-\gamma_2\vert^2)$ when $Y\rightarrow Y_{\text{EP}}$ is shown in Appendix ~\ref{AppendixV}. $C_4/\vert a_1+a_2+3Y\vert\approx0.447$ at EP when $k=10$. Equation (\ref{R_const}) shows clearly that when $Y\rightarrow Y_{\text{EP}}$, for large $x$, $R$ may be very large.

\section{\label{sec_flux}The influence of using TSP to calculate $G_{max}$}

\begin{figure}[htbp]\label{fig_flux}
\centering\includegraphics[width=0.7\textwidth]{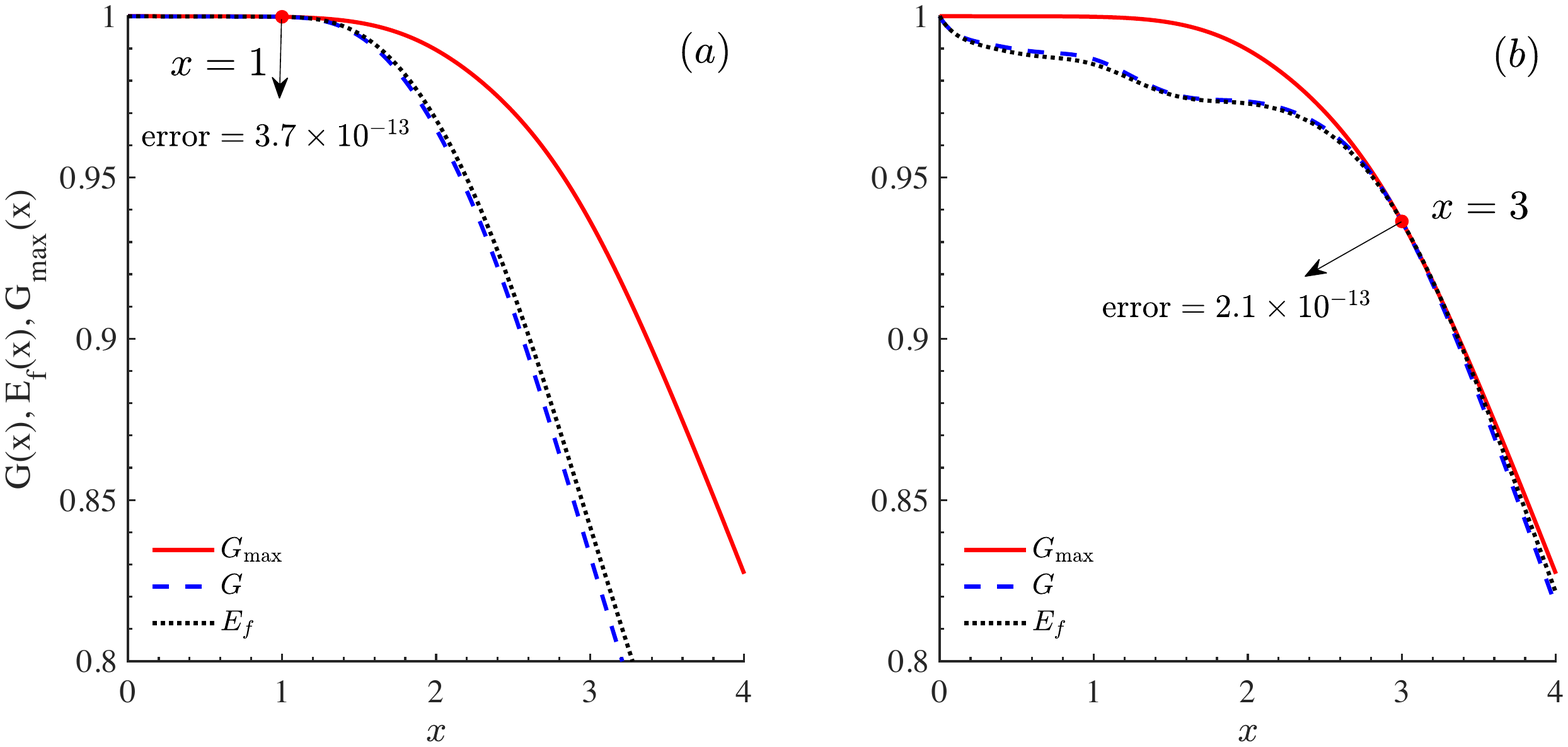}
\caption{(color online) Comparisons between the attenuated TSP $G(x)$ (dashed line) and the attenuation of energy flux $E_f(x)$ (dot line) when the least attenuated source: (a) $\mathbf{q}_{least}(x_f=1)$;  (b) $\mathbf{q}_{least}(x_f=3)$ are incident, respectively. $G_{max}$ (solid line) is also plotted. $Y=1+4\text{i}$, $k=10$.}
\end{figure}

All the above results are based on the attenuation of TSP defined in Eq. (\ref{eq_def_power}). However, we usually use energy flux to describe the attenuation of energy in waveguides. To verify whether this difference will have important influence on our conclusions, we re-plot the $G_{max}(x)$ in Fig.~\ref{fig_flux} (a) and (b) by using the parameters $Y=1+4\text{i}$ and $k=10$ as shown in Fig. ~\ref{fig_Gmax}(c). Using the same parameters, we plot $G(x)$ when $\mathbf{q}_{least}(x_f=1)$ and $\mathbf{q}_{least}(x_f=3)$ are incident respectively. In the same figure, we also plot the attenuation of the energy flux defined as
\begin{equation}\label{eq_flux}
E_{f}(x)=\frac{\text{Im}\big(\int_0^1{p^*\frac{\partial p}{\partial x}}\, \mathrm{d}y\big)}{\text{Im}\big(\int_0^1{\big(p^*\frac{\partial p}{\partial x}\big)\big|_{x=0}}\, \mathrm{d}y\big)}, 
\end{equation}   
by incidence of  the same least attenuated sources. It is shown that $E_f(x)$ is very close to $G(x)$. At $x=1$ and $x=3$, the differences between $E_f(x)$ and $G(x)$ are of the order $O(10^{-13})$ and therefore could be negligible. We have implemented many comparisons for different $\mathbf{q}_{least}(x_f)$ to justify the above results. We then confidently conclude that all our conclusions about $G_{max}$ obtained from $G(x)$ defined in Eq. (\ref{eq_def_power}) are valid for energy flux defined in Eq. (\ref{eq_flux}).  

\section{\label{conclusion}Conclusions}

We have studied the spatial transient behavior of the least attenuated total sound power $G_{max}(x)$ described by the largest singular value of exp($\text{i}x\mathsf{L}$) in a waveguide with lossy admittance boundary conditions which can be described by a non-Hermitian and non-normal operator $\mathsf{L}$. $G_{max}(x)$ is not curves of total sound power attenuation along wave propagation when sources input as usually illustrated in Acoustic literature. It is the envelop of them. $G_{max}(x)$ can be roughly divided into three regions: (1) almost non-decayed transient region in which $G_{max}$ is almost non-decayed although all modes attenuated exponentially. This almost non-decayed transient behavior can be realized for any complex value admittance, even when it is close to the EPs, the spectral singularities of non-Hermitian operators, at which a pair of adjacent modes achieve maximum attenuation rates predicted by Cremer optimum criterion. (2) damped-oscillation transient region in which $G_{max}(x)$ may be analogous to the displacement of a damped harmonic oscillator in underdamping or overdamping. (3) asymptotic region.

We have conducted detail studies in the differences between $G_{max}$ and $M_{max}$ which describes the attenuation of the least attenuated mode. We have shown that the departures of $G_{max}$ from $M_{max}$ depend on $x$ and the lossy admittance boundary conditions. The maximum departure, defined as non-normality indicator, tends to be infinite when the admittance is close to the EPs. Only in the asymptotic region, $G_{max}(x)$ decays with the same rates as $M_{max}(x)$. However, $G_{max}(x)$ and $M_{max}(x)$ still differ by a constant. This constant tends to be infinite when the admittance is close to the EPs. 

We have shown that for each given $x$, the corresponding $G_{max}(x)$ is achieved by a different source. When this source inputs at the start of the waveguide $x=0$, the sound field can avoid the lossy admittance boundary, like a half of a Gaussian beam, in the almost non-decayed transient region.

Our analysis based on singular-value decomposition and a two-by-two toy model is generic and capture not only the asymptotic behavior governed only by eigenvalues, but the spatial transient behavior in which the eigenvalues may be misleading and non-normality plays an important role. We believe that our analysis approach and conclusions would be interesting for other lossy acoustic systems and might shed new light on other lossy physical systems.

\begin{appendices}

%
\renewcommand{\theequation}{A-\arabic{equation}}
\setcounter{equation}{0}
\section{\label{AppendixI} Derivation of Eqs. (\ref{eq_matrix_x}), (\ref{toy1}) and (\ref{toy_matrix})}

We consider the semi-infinite, two-dimensional ($2$D) waveguide as defined by the configuration depicted in Fig.~\ref{fig2} and the dimensionless Helmholtz equation (\ref{eq1}) and boundary conditions (\ref{BC}). All length variables are divided by the waveguide width $h$. Time dependence $\text{e}^{-\text{i}\omega t}$ is omitted in the followings. 

To derive Eq. (\ref{eq_matrix_x}), we have used a spectral collocation method based on Chebyshev polynomials to discretize sound pressure $p(x,y)$ along $y$, $\mathbf{p}^T=[p_1, p_2, \cdots, p_n, \cdots, p_N]^T=[p(y_1), p(y_2), \cdots, p(y_n), \cdots, p(y_N)]^T$, where $\mathbf{p}$ is a column vector, $``{}^T"$ refers to the transpose, $N$ refers to truncation number, the interpolation points $y_n=\cos[(n-1)\pi/(N-1)]$, $n=1,\;\cdots\; N$. The differential operator $\partial^2/\partial y^2$ is approximated by $\mathsf{D_2}$ calculated by the MATLAB program \textit{chebdif}. The Eq. (\ref{eq1}) is then written as 
\begin{equation}
\frac{\partial^2\mathbf{p}}{\partial x^2}=-(\mathsf{D_2}+k^2\mathsf{I})\mathbf{p}.
\end{equation}
In an infinite and uniform (range independent) waveguide, $\mathbf{p}$ can be decomposed as right- and left-going waves, $\mathbf{p}=\mathbf{p}^++\mathbf{p}^-$ with
\begin{equation}\nonumber
\frac{d\mathbf{p}^+}{dx}=\text{i}\mathsf{L}\mathbf{p}^+,\;\; \frac{d\mathbf{p}^-}{dx}=-\text{i}\mathsf{L}\mathbf{p}^-,
\end{equation}
where $\mathsf{L}=\sqrt{\mathsf{D_2}+k^2\mathsf{I}}$ is a matrix operator,  $\mathsf{I}$ refers to identity matrix. We consider only right-going wave from the source at $x=0$ and suppress the ``+", obtain
\begin{equation}\nonumber
\frac{d\mathbf{p}}{dx}=\text{i}\mathsf{L}\mathbf{p}.
\end{equation}

To derive Eqs. (\ref{toy1}) and (\ref{toy_matrix}), we project Eq. (\ref{eq1}) onto the corresponding rigid normalized eigenfunctions 
\begin{equation}
\psi_n(y)=\sqrt{\epsilon_n}\cos(\alpha_ny),\;\; \text{with} \;\; \left\{\begin{aligned} & \epsilon_0=1,\;\epsilon_n=2, \; \text{for} \; n\ge1,\\ & \alpha_n=n\pi, \end{aligned}\right.\;\; \text{and}\;\;\int_0^1\psi_m(y)\psi_n(y)dy=\left\{\begin{aligned} 1,\; m=n,\\0, \; m\ne n, \end{aligned}\right.
\end{equation}
as
\begin{equation}\nonumber
\int_0^1\bm{\psi}\left(\frac{\partial^2 p}{\partial x^2}+\frac{\partial^2 p}{\partial y^2}+k^2p\right)dy=0,
\end{equation}
where $\bm{\psi}=[\psi_0, \psi_1, \cdots, \psi_N]^T$, $``{}^T"$ refers to the transpose, $N$ the truncation number of vector $\bm{\psi}$. By using the expansion (\ref{eq_expand}), we obtain
\begin{equation}\nonumber
\frac{\partial^2\mathbf{p}}{\partial x^2}+(k^2\mathbf{I}-\bm{\alpha}+Y\bm{\psi}(y=1)\bm{\psi}^T(y=1))\mathbf{p}=0,
\end{equation}
where the elements on the main diagonal of the diagonal matrix $\bm{\alpha}$ are $\alpha_n^2$, $n=0, 1,\cdots, N$. The above equation depends only on $x$, it is then rewritten as
\begin{equation}
\frac{d^2\mathbf{p}}{dx^2}=-\mathsf{L}_1\mathbf{p},\;\; \text{with}, \;\; \mathsf{L}_1=k^2\mathbf{I}-\bm{\alpha}+Y\bm{\psi}(y=1)\bm{\psi}^T(y=1).
\end{equation}
Similar to the derivation above, we obtain Eq. (\ref{toy1})
\begin{equation}
\frac{d\mathbf{p}}{dx}=\text{i}\mathsf{L}\mathbf{p}
\end{equation}
where $\mathsf{L}=\sqrt{\mathsf{L}_1}$. If we truncate matrix $\mathsf{L}_1$ to $N=1$, Eq. (\ref{toy_matrix}) is obtained.

To calculate the elements of $\mathsf{L}$, we first calculate the eigenvalues and their corresponding normalized eigenvectors of matrix $\mathsf{L}_1$ defined in Eq. (\ref{toy_matrix}), which are
\begin{eqnarray}\label{eig_lambda}
\gamma_{1} &=& \frac{1}{2}\left(a_1+a_2+3Y+\sqrt{\left(\pi^2-Y\right)^2+8Y^2}\right),\\
\gamma_{2} &=& \frac{1}{2}\left(a_1+a_2+3Y-\sqrt{\left(\pi^2-Y\right)^2+8Y^2}\right),\nonumber
\end{eqnarray}
\begin{equation}\label{eigv_x}
\bm{X}_{1,2}=\frac{1}{\Lambda_{1,2}}\left[-\sqrt{2}Y, \;\;\gamma_{1,2}-(a_1+Y)\right]^T,
\end{equation}
where the normalization constants $\Lambda_{1,2}$ are defined as
\begin{equation}\label{normal_x}
 \Lambda_{1,2}=\sqrt{2\vert Y\vert^2+\vert\gamma_{1,2}-(a_1+Y)\vert^2}, \;\; \text{such that}\;\;  \bm{X}_{n}{^\dagger}\bm{X}_{n}=1,\; n=1,2
\end{equation}
where $``{}^\dagger"$ refers to the complex conjugate transpose.
We can show that 
\begin{equation}
\bm{X}_{m}{^\dagger}\bm{X}_{n}\ne 0,\;\; m\ne n
\end{equation}
i.e., eigenvectors are not mutually orthogonal. This non-orthogonality is a general property of a non-normal matrix. 
Besides, we can find the basis that is biorthogonal to $\mathbf{X}_1$ and $\mathbf{X}_2$. In our case, since the matrix $\mathsf{L}_1$ is complex symmetric
($\mathsf{L_1}^T = \mathsf{L}_1$), this biorthognal basis is simply $\mathbf{X}_1^*$ and $\mathbf{X}_2^*$ such that
$\bm{X}_{m}{^T}\bm{X}_{n}=0$, $m\ne n$.

By using the bi-orthogonal eigenfunctions $\bm{X}_{1, 2}$ of matrix $\mathsf{L}_1$, $\mathsf{L}$ can be expressed as 
\begin{equation}\label{L_matrix}
\mathsf{L}= \left[\bm{X}_{1}, \bm{X}_{2}\right]\begin{bmatrix}k_1 & 0\\ 0 & k_2\end{bmatrix}\left[\bm{X}_{1}, \bm{X}_{2}\right]^{-1}=\begin{bmatrix} 
\mathsf{L}_{11} & \mathsf{L}_{12}\\ \mathsf{L}_{21} & \mathsf{L}_{22}\end{bmatrix},
\end{equation}
where $k_1=\sqrt{\gamma_1}$ and $k_2=\sqrt{\gamma_2}$ are the eigenvalues of matrix $\mathsf{L}$ or axial wavenumbers of wave $\mathbf{p}(x)$. $\mathsf{L}_{11}$, $\mathsf{L}_{12}$, $\mathsf{L}_{21}$, and $\mathsf{L}_{22}$ are then obtained as follows
\begin{eqnarray}\label{L_element}
\mathsf{L}_{11} &=&  \frac{1}{2}(k_1+k_2)+\frac{1}{2}\frac{k_1-k_2}{\gamma_1-\gamma_2}\left(\pi^2-Y\right),\\\nonumber
\mathsf{L}_{12} &=& -\frac{k_1-k_2}{\gamma_1-\gamma_2}\sqrt{2}Y,\\\nonumber
\mathsf{L}_{21} &=& -\frac{k_1-k_2}{\gamma_1-\gamma_2}\sqrt{2}Y,\\\nonumber
\mathsf{L}_{22} &=&  \frac{1}{2}(k_1+k_2)-\frac{1}{2}\frac{k_1-k_2}{\gamma_1-\gamma_2}\left(\pi^2-Y\right).\\\nonumber
\end{eqnarray}

\renewcommand{\theequation}{B-\arabic{equation}}
\setcounter{equation}{0}
\section{\label{AppendixII} Validity of the two-by-two toy model as a true passive system} 

To assure the truncation of the expansion (\ref{eq_expand}) to $N=1$ represents a true passive sound pressure field in a waveguide with lossy impedance boundary conditions, we must choose the imaginary parts of eigenvalues $k_1$ and $k_2$ to be non-negative, and the axial variation of TSP, $dW(x)/dx<0$, where 
\begin{equation}\label{dwdx}
\frac{dW(x)}{dx}=\frac{d(\mathbf{q}^\dagger\mathbf{q})}{dx}=\mathbf{q}^\dagger\mathsf{L}_2\mathbf{q},
\end{equation}  
where $\mathsf{L}_2=\text{i}(\mathsf{L'}-\mathsf{L'}^\dagger)$. In the derivation of Eq. (\ref{dwdx}), we have used the expansion Eq. (\ref{eq_expand}) and the orthogonal properties of $\psi_n$. It is easy to verify that $\mathsf{L}_2$ is Hermitian so that its eigenvalues $\gamma_{\pm}=-\text{Im}(k_1+k_2)\pm\sqrt{[\text{Im}(k_1-k_2)]^2+\vert C_1\vert^2}$ are real. To assure $dW(x)/dx<0$, $\mathsf{L}_2$ has to be negative semidefinite, i.e., $\gamma_\pm$ are nonpositive. This is equivalent to the condition that 
\begin{equation}
4\text{Im}(k_1)\text{Im}(k_2)>\vert C_1\vert^2.
\end{equation} 
We have numerically verified this condition over complex $Y$ plane. Note that Wiersig~\cite{wiersig} has derived a similar condition for a two-by-two non-Hermitian passive system.

\renewcommand{\theequation}{C-\arabic{equation}}
\setcounter{equation}{0}
\section{\label{AppendixIII} Derivations of Eq. (\ref{toy11})}

To analytically calculate the least attenuated TSP $G_{max}(x)$, we transform the matrix $\mathsf{L}$ to a triangle matrix by a change-of-base matrix $\mathsf{T}=[\bm{X}_{1}, \bm{Y}_{2}]$, $\bm{Y}_{2}$ is
\begin{equation}
\bm{Y}_{2}=\frac{1}{\Lambda_{1}}\left[\gamma_{1}^*-(a_1+Y)^*, \;\; \sqrt{2}Y^*\right]^T,
\end{equation}
where ``${}^*$" refers to complex conjugate. $\bm{X}_{1}$ and $\bm{Y}_{2}$ verify the standard orthonormal relation $\bm{X}_{1}{^\dagger}\bm{Y}_{2}=0$, and normalization $ \bm{Y}_{2}{^\dagger}\bm{Y}_{2}=1$, therefore, $\mathsf{T}$ is a unitary matrix. This unitary property makes the transformation $\mathbf{q}=\mathsf{T}^{-1}\mathbf{p}$ to conserve the TSP. 

By using the unitary transformation $\mathbf{q}=\mathsf{T}^{-1}\mathbf{p}$, Eq. (\ref{toy1}) can be written as
\begin{equation}
\frac{\mathrm{d}{\mathbf{q}}}{\mathrm{d}x}=\text{i}{\mathsf{L'}}\mathbf{q}, 
\end{equation}
where
\begin{equation}\label{toy11a}
\mathsf{L}'=\mathsf{T}^{-1}\mathsf{L}\mathsf{T}=\begin{bmatrix}k_1 & C_1 \\ 0 & k_2 \end{bmatrix},\;\;\text{and}\;\; C_1=-\text{i}\frac{\sqrt{2}\pi^2[\pi^2-Y-(\gamma_1-\gamma_2)]^*}{\Lambda_1^2}\frac{k_1-k_2}{\gamma_1-\gamma_2}\text{Im}(Y).
\end{equation}

\renewcommand{\theequation}{D-\arabic{equation}}
\setcounter{equation}{0}
\section{\label{AppendixIV} Derivations of Eq. (\ref{Rx_approx})}

In this Appendix, we derive the approximation of Eq. (\ref{Rx_approx}) under the assumption of $x\ll 1$ such that $k_\mathbb{R}x\ll 1$ and $k_\mathbb{I}x\ll 1$.   

Based on Taylor series, we first analyze $e^{-2k_{\mathbb{I}}x}-2\cos(k_\mathbb{R}x)e^{-k_{\mathbb{I}}x}+1$ by keeping to $(k_\mathbb{R}x)^2$ and $(k_\mathbb{I}x)^2$
\begin{eqnarray}
e^{-2k_{\mathbb{I}}x}-2\cos(k_\mathbb{R}x)e^{-k_{\mathbb{I}}x}+1 &\approx& 1-2k_\mathbb{I}x+2k_\mathbb{I}^2x^2-2(1-\frac{1}{2}k_\mathbb{R}^2x^2)(1-k_\mathbb{I}x+\frac{1}{2}k_\mathbb{I}^2x^2)+1\nonumber\\
&=& (k_\mathbb{R}^2+k_\mathbb{I}^2)x^2.
\end{eqnarray}
Then
\begin{equation}
\left[e^{-2k_{\mathbb{I}}x}-2\cos(k_\mathbb{R}x)e^{-k_{\mathbb{I}}x}+1\right]P_t\approx 4C_2\frac{k_\mathbb{R}^2+k_\mathbb{I}^2}{\vert\gamma_1-\gamma_2\vert^2} x^2,
\end{equation}
where we have used the $P_t$ expression in Eq. (\ref{G_analy1}). For ease of presentation, we assume $A=(k_\mathbb{R}^2+k_\mathbb{I}^2)/\vert\gamma_1-\gamma_2\vert^2$ in this Appendix. Therefore, 
\begin{equation}
\text{tr}=2-2k_\mathbb{I}x+\left(2k_\mathbb{I}^2+C_4A\right)x^2,
\end{equation}
where $C_4=4C_2C_3\text{Im}(Y)^2$.

In the similar way, we have
\begin{eqnarray}
\sqrt{\text{tr}^2-4e^{-2\text{Im}(k_{\mathbb{I}})x}} &\approx& \sqrt{\left(2k_\mathbb{I}^2+C_4A\right)^2x^2-4k_\mathbb{I}\left(2k_\mathbb{I}^2+C_4A\right)x+4\left(k_\mathbb{I}^2+C_4A\right)}\;x\nonumber\\
&\approx& 2\sqrt{k_\mathbb{I}^2+C_4A}\left\{1+\frac{\left(2k_\mathbb{I}^2+C_4A\right)\left[\left(2k_\mathbb{I}^2+C_4A\right)x-4k_\mathbb{I}\right]x}{8\left(k_\mathbb{I}^2+C_4A\right)}\right\}x\\
&=&2\sqrt{k_\mathbb{I}^2+C_4A}x-\frac{k_\mathbb{I}\left(2k_\mathbb{I}^2+C_4A\right)}{\sqrt{k_\mathbb{I}^2+C_4A}}x^2,\nonumber
\end{eqnarray}
where we have used Taylor series to approximate the square root.
By using Eq. (\ref{tbt_MR}), we obtain the approximation of $R(x,Y)$ as Eq. (\ref{Rx_approx}). 

The coefficients of $x$ and $x^2$ in Eq. (\ref{Rx_approx}) are both non-negative. Since $C_4\ge 0$, we can easily verify the coefficient of $x$ is always non-negative. For the coefficient of $x^2$, since
\begin{equation}\nonumber
2k_\mathbb{I}^2+C_4A-\frac{k_\mathbb{I}\left(2k_\mathbb{I}^2+C_4A\right)}{\sqrt{k_\mathbb{I}^2+C_4A}}>2k_\mathbb{I}^2+C_4A-2k_\mathbb{I}\sqrt{k_\mathbb{I}^2+C_4A},
\end{equation} 
To verify the right hand side of above equation is greater than $0$, we need only to show
\begin{equation}\nonumber
2k_\mathbb{I}^2+C_4A>2k_\mathbb{I}\sqrt{k_\mathbb{I}^2+C_4A},
\end{equation} 
by just squaring both sides of this equation because each term is non-negative.

\renewcommand{\theequation}{E-\arabic{equation}}
\setcounter{equation}{0}
\section{\label{AppendixV} Approximation of $(k_\mathbb{R}^2+k_\mathbb{I}^2)/\vert\gamma_1-\gamma_2\vert^2$ when $Y\rightarrow Y_{\text{EP}}$}

We first analyze the behavior of $k_{\mathbb{R}}$ (or $k_{\mathbb{I}}$) when $Y\rightarrow Y_{\text{EP}}$. For ease of presentation, we assume that $\epsilon=Y-Y_{\text{EP}}$, $C_5=Y-Y_{\text{EP}}^*$, $C_6=a_1+a_2+3Y$, by using Eqs. (\ref{eig_lambda}) and (\ref{gamma_YEP}), we have
\begin{equation}\nonumber
k_1=\sqrt{\gamma_1}=\frac{1}{\sqrt{2}}\sqrt{C_6+\sqrt{C_5\epsilon}}.
\end{equation}
When $Y\rightarrow Y_{\text{EP}}$, $\epsilon\rightarrow 0$, we use Taylor series expansion to obtain
\begin{equation}\label{k1approx}
k_1\approx\frac{\sqrt{C_6}}{\sqrt{2}}\left(1+\frac{1}{2}\frac{\sqrt{C_5\epsilon}}{C_6}\right)=\frac{1}{\sqrt{2}}\sqrt{a_1+a_2+3Y}+\frac{1}{2\sqrt{2}}\sqrt{\frac{(Y-Y_{\text{EP}}^*)(Y-Y_{\text{EP}})}{a_1+a_2+3Y}}.
\end{equation}
Again, we can obtain the approximation of $k_2$ when $Y\rightarrow Y_{\text{EP}}$. Therefore,
\begin{equation}
\Delta k=k_1-k_2\approx\frac{1}{\sqrt{2}}\sqrt{\frac{C_5\epsilon}{C_6}}\nonumber.
\end{equation}
$k_{\mathbb{R}}$ and $k_{\mathbb{I}}$ are then obtained as
\begin{eqnarray}\label{kRkI}
k_{\mathbb{R}} &=& \text{Re}(\Delta k)=\frac{1}{\sqrt{2}}\sqrt{\frac{\vert C_5\vert\vert\epsilon\vert}{\vert C_6\vert}}\cos\frac{\theta_k}{2},\\
k_{\mathbb{I}} &=& \text{Im}(\Delta k)=\frac{1}{\sqrt{2}}\sqrt{\frac{\vert C_5\vert\vert\epsilon\vert}{\vert C_6\vert}}\sin\frac{\theta_k}{2},\label{kRkI1}
\end{eqnarray}
where $\theta_k=\theta_{C_5}+\theta_{\epsilon}-\theta_{C_6}$, $\theta_{C_5}$, $\theta_{\epsilon}$, and $\theta_{C_6}$ are the arguments of $C_5$, $\epsilon$, and $C_6$ respectively. $(k_\mathbb{R}^2+k_\mathbb{I}^2)/\vert\gamma_1-\gamma_2\vert^2$ is then obtained by using Eqs. (\ref{kRkI}), (\ref{kRkI1}), and (\ref{gamma_YEP}), when $Y\rightarrow Y_{\text{EP}}$
\begin{equation}
\frac{k_\mathbb{R}^2+k_\mathbb{I}^2}{\vert\gamma_1-\gamma_2\vert^2}=\frac{1}{2\vert C_6\vert}=\frac{1}{2\vert a_1+a_2+3Y\vert}.
\end{equation}

\end{appendices}

\section*{Acknowledgements}
The authors Wei Guo and Juan Liu gratefully acknowledge support by China Scholarship Council (CSC) and LMAc.

\bibliographystyle{elsarticle-num} 
\bibliography{Bib_anomalous.bib}
\end{document}